\documentclass[twocolumn,pre,aps,10pt,showpacs,superscriptaddress]{revtex4-1}

\usepackage{graphicx}
\usepackage{amssymb,amsfonts,amsmath}
\usepackage{epstopdf}

    \usepackage{multirow}
    
\begin{document}

\title{Characterizing protein crystal contacts and their role in crystallization: rubredoxin as a case study.}

\author{Diana Fusco}
\affiliation{Program in Computational Biology and Bioinformatics, Duke University, Durham, NC 27708}

\author{Jeffrey J. Headd}
\affiliation{Department of Biochemistry, Duke University, Durham, NC 27708}

\author{Alfonso De Simone}
\affiliation{Division of Molecular Biosciences, Imperial College London, South Kensington SW7 2AZ, UK}

\author{Jun Wang}
\affiliation{Department of Chemistry, Duke University, Durham, NC 27708}

\author{Patrick Charbonneau}
\affiliation{Program in Computational Biology and Bioinformatics, Duke University, Durham, NC 27708}
\affiliation{Department of Chemistry, Duke University, Durham, NC 27708}
\affiliation{Department of Physics, Duke University, Durham, NC 27708}

\begin{abstract}
The fields of structural biology and soft matter have independently sought out fundamental principles to rationalize protein crystallization. Yet the conceptual differences and the limited overlap between the two disciplines have thus far prevented a comprehensive understanding of the phenomenon to emerge. We conduct a computational study of proteins from the rubredoxin family that bridges the two fields.
Using atomistic simulations, we characterize their crystal contacts, and accordingly parameterize patchy particle models. Comparing the phase diagrams of these schematic models with experimental results enables us to critically examine the assumptions behind the two approaches. The study also reveals features of protein-protein interactions that can be leveraged to crystallize proteins more generally.
\end{abstract}
 
\maketitle

\section{Introduction}

%Crystallography may be the gold standard of protein structure determination, but obtaining the necessary high-quality crystals is akin to prospecting for the precious mineral~\cite{mcpherson:1999}. 
In spite of the recent advances in NMR techniques~\cite{Grzesiek2009}, 
X-ray and neutron diffraction crystallography remain the methods of choice for high-precision protein structure determination. 
Sophisticated screening methods and the parallel testing of several different crystallization conditions have significantly increased the number of deposited protein structures and complexes~\cite{mcpherson:1999}.
Yet the lack of systematic ways to crystallize proteins still limits the timely and cost-effective use of crystallography. This experimental bottleneck notably constrains our understanding of certain biochemical mechanisms and our ability to design better drugs and biomaterials~\cite{mcpherson:1999,chayen:2008,Chayen2009,blundell:2002,kuhn:2002,blundell:2004,tickle:2004,kitchen:2004,congreve:2005,huebsch:2009}. Developing a more quantitative characterization of protein crystallization is therefore fundamental to advance both biological and bio-inspired research.
%The challenges of designing better drugs~\cite{blundell:2002,kuhn:2002,blundell:2004,tickle:2004,kitchen:2004,congreve:2005} and biomaterials~\cite{huebsch:2009}, which partially
%rest on determining the target proteins' precise structure, are thus harder to meet. Incomplete structural information for a number of biologically important proteins, such as the ribosome~\cite{schmeing:2009,Dunkle2011}, also limits our understanding of their function. 

From a physical viewpoint, protein crystallization should follow from a detailed description of protein-protein interactions~\cite{mcpherson:1999,derewenda:2010,anderson:2006,saridakis:2009}. 
%But although the underlying forces, i.e. hydrogen bonding, van der Waals and hydrophobic interactions, electrostatics, are individually fairly well characterized~\cite{israelachvili:1991,Chandler2005,gunton:2007,granick:2008,Berne2009}, their collective contribution to protein assembly is much less clear~\cite{vekilov:2002}. 
In contrast to the interactions that drive protein complex formation and protein-target association, which are on average stronger and evolutionarily tuned to be selective, the interactions that drive crystallization are thought to be \textit{non-specific}\footnote[2]{
The meaning of \textit{non-specific protein-protein interactions} depends on the disciplinary context. In chemistry, it distinguishes certain attraction forces from others, although the classification of the various physical mechanisms in not unambiguous~\cite[(\S~18.8)]{israelachvili:1991}. In biophysics, the distinction between specific and non-specific interactions typically relies on the existence of an energy gap that clearly divides a single, strong (specific) interaction from the other (non-specific) ones~\cite{janin:1995,johnson:2011}. In molecular biology, specific interactions are deemed responsible for the stoichiometric recognition of a given target, while non-specific interactions arise from the promiscuous yet non-biologically relevant association of molecules~\cite{janin:1995b,carugo:1997,wilkinson:2004,zhuang:2011}. Specific interactions have thus been evolutionarily tuned to be free-energetically strong and geometrically oriented, while non-specific attractions have not. This general weakness, however, may itself have evolved so as to prevent pathological aggregation~\cite{janin:1995,doye:2004}. Note that although these three definitions are not necessarily orthogonal to one another, we here specifically aim to clarify the last one. When applied to crystal contacts it has indeed been used to suggest that these biologically non-functional interactions are in many ways indistinguishable from interfaces obtained by randomly bringing two proteins together~\cite{janin:1995}.}.
%They are generally weak, biologically non-functional and have long been thought to be indistinguishable from random surface patches~\cite{janin:1995b,carugo:1997,zhu:2006}}.
%In the field of structural biology, protein-protein interactions have historically been divided in two categories: \emph{specific} interactions that are responsible for ``functionally relevant assemblies", such as protein complex formation or for protein-target association, and \emph{nonspecific} interactions that are ``short-lived and insignificant", randomly occurring between two protein coils~\cite{bahadur:2004}. 
%The weak interactions that result in crystal contacts, i.e., the regions of the protein surface that lie within 0.4nm from another chain in the crystal phase~\cite{Zhang:1995}, are commonly cited as examples of nonspecific interactions. They have indeed long been thought to be indistinguishable from randomly selected surface patches~\cite{janin:1995b,carugo:1997,bahadur:2004}.
Two recent studies, however, present crystal contacts under a more probing light. First, Cie\'slik and Derewenda found that crystal contacts are enriched for glycine and small hydrophobic residues, and depleted in large polar residues with high side-chain entropy, such as lysine and glutamic acids~\cite{Derewenda2009}. Second, mining a database recording the output of hundreds of crystallization experiments, Price \emph{et al.} found that proteins with a large fraction of glycine and alanine on their surface are more likely to have been successfully crystallized~\cite{price:2009}. These observations provide immediate statistical support for the surface entropy reduction (SER) mutagenesis strategy, which recommends replacing high-entropy surface residues with alanine to facilitate crystal formation~\cite{derewenda:2004,derewenda:2010,Lanci2012}. More fundamentally, these studies also 
suggest that crystal contacts correspond to non-randomly selected regions of the protein surface. To some degree, it should thus be possible to control these weak yet directional, i.e. \textit{patchy}, protein-protein interactions by tuning the solution conditions or by mutating certain surface residues.  A key missing insight to developing crystallization strategies is thus understanding the context in which these interactions can result in regular protein assembly.

The soft matter viewpoint on particle interactions presents a possible answer to this challenge.
The observation that short-range isotropic attraction between particles results in a gas-liquid critical point that lies below the crystal solubility regime~\cite{gast:1983,Hagen1994}, in particular, offers a first analogy for the solution behavior of proteins~\cite{lomakin:1996,rosenbaum:1996,chayen:2005}. In these model systems successful crystallization can most easily be achieved in the region between the solubility line, above which the system does not aggregate because the disperse phase is stable, and the critical point, below which the system typically forms ``amorphous'' materials that are useless for crystallography~\cite{Lu2008}.  
Though appealing in their simplicity, isotropic description of protein-protein interactions between proteins are, however, too simplistic~\cite{haas:1999,curtis:2001,lomakin:1996,lomakin:1999,mcmanus:2007}. 
Partly in response to this difficulty, a broad array of schematic models with anisotropic, directional attraction, i.e., \textit{patchy} models, have been developed~\cite{gogelein:2008,Bianchi2011}.

Although both the structural biology and the soft matter fields target the same problem, a large gap between the two research lines remains to be filled before synergistic experimental guidance can be provided. In particular,  although anisotropy plays a key role in physical models for protein crystallization~\cite{Kern2003,charbonneau:2007b,foffi:2007}, little characterization of the directional interaction between proteins at crystal contacts has been done~\cite{Pellicane2008}, leaving most of the physical assumptions behind patchy models untested. Can these models explain the results of crystallographic experiments if they are parameterized using actual protein-protein interactions?  If yes, the relation between the resulting phase diagrams and protein-protein interactions should allow one to rationally alter these interactions, in order to control protein crystal assembly.

%Despite the convergence between the structural biology and the soft matter research lines on the relevance of interaction anisotropy for protein crystallization, a large conceptual gap remains to be filled before reliable experimental guidance can be provided. In particular,  although anisotropy plays a key role in physical models for protein crystallization~\cite{Kern2003,charbonneau:2007b,foffi:2007}, little characterization of the actual interaction between proteins at crystal contacts has been done~\cite{Pellicane2008}, leaving most of the physical assumptions behind patchy models untested. Are these models able to explain the experimental results if they are parameterized using actual protein-protein interactions?  

In this article, we answer this question for simple proteins of the rubredoxin family, using a hybrid atomistic and schematic simulation approach. Classical atomistic simulations characterize the differences and similarities in the crystal contact interactions of three closely related small globular proteins from the rubredoxin family: (a) the wild-type from \textit{Pyrococcus furious} (wt-RbPf, PDB code: 1BRF)~\cite{Bau1998}, (b) its W3Y/I23V/L32I mutant (mut-RbPf, PDB code: 1IU5)~\cite{Chatake2004}, and (c) the W4L/R5S mutant from \textit{Pyrococcus abyssi} (mut-RbPa, PDB code: 1YK4)~\cite{Bonish2005}. Through a comparative analysis, we identify the molecular basis of these protein-protein interactions, and parameterize patchy models whose phase diagrams are then compared with experimental crystallization conditions. The validity of this strategy is supported by the recent success of multiscale descriptions of protein aggregation~\cite{DeSimone2012}. By showing that the models and the experimental results agree fairly well, we find that increasing the solution temperature may sometimes produce better crystallization conditions. We also suggest ways to improve SER and sketch a framework for developing physically representative patchy models of proteins.

The plan of this paper is as follows. 
In Section~\ref{sect:model}, we describe the atomistic and schematic models as well as the corresponding molecular dynamics (MD) and Monte Carlo methodologies. In Section~\ref{sect:results}, we report the MD potential of mean force (PMF) analysis for each protein and the phase diagrams of the corresponding schematic models. We then compare these phase diagrams to experimental crystallization conditions, which help understand the role of salt in rubredoxin crystallization. Section~\ref{sect:discussion} discusses how our findings illuminate the SER method and the patchy particle models of proteins. Section~\ref{sect:conclusion} summarizes our conclusions and discusses possible future research directions.

\section{Models and Simulation Methods}
\label{sect:model}
Hypethermophilic rubredoxins are an excellent model system for the computational study we present here. First, their core is characterized by more hydrogen bonding and electrostatic interactions than those of their mesophilic counterparts, which reduces their conformational flexibility and justifies the use of relatively short molecular dynamics simulations to capture the relevant protein dynamics. Second, these proteins have a tight hydrophobic core packing, which motivates the hard sphere analogy. Third, they are structurally stable over a wide temperature range, which makes the PMF (and thus the phase diagram) predictions reasonably transferable to temperatures beyond those used in the molecular dynamics simulations~\cite{Dalhus:2002}.

\subsection{Iron site model and Molecular Dynamics simulations.}

Molecular dynamics (MD) simulations are performed with the Gromacs package~\cite{gromacs} using the Amber99sb forcefield~\cite{Hornack2006} and explicit TIP4P water with Ewald summation~\cite{tip4p}. The full list of simulation parameters is reported in Table~\ref{table:MD}. 

\begin{table}[h]
\small
\centering
\caption{MD simulations parameters}\label{table:MD}
\begin{tabular}{c c}
\hline
\textbf{Parameter}&\textbf{Value}\\
\hline
Forcefield&Amber99sb~\cite{Hornack2006}\\
Water model&Tip4pEW~\cite{tip4p}\\
Ions&Aqvist~\cite{aqvist:1990}\\
Temperature&300 K\\
Temperature control&Nos\'{e}-Hoover thermostat~\cite{nose:1984,hoover:1985}\\
Pressure&1 bar\\
Pressure control&Parrinello-Rahman barostat~\cite{parrinello:1981}\\
Box dimension&6 nm$\times$6 nm$\times$12 nm\\
Periodic boundary conditions&xyz\\
Electrostatic method&PME~\cite{essmann:1995}\\
Coulomb radius&1.4 nm\\
Van der Waals method&Cut-off\\
Van der Waals radius&1.4 nm\\
Integration step&2 fs\\
Constraint algorithm&Lincs~\cite{lincs}\\
Energy minimization&steepest-descent\\
Salt concentration&45 mM/3 M of NaCl\\
Spring constant&5000 kJ/(mol nm$^2$)\\
\hline
\end{tabular}
\end{table}

\begin{figure}[htb]
\centering
\includegraphics[width=0.2\textwidth]{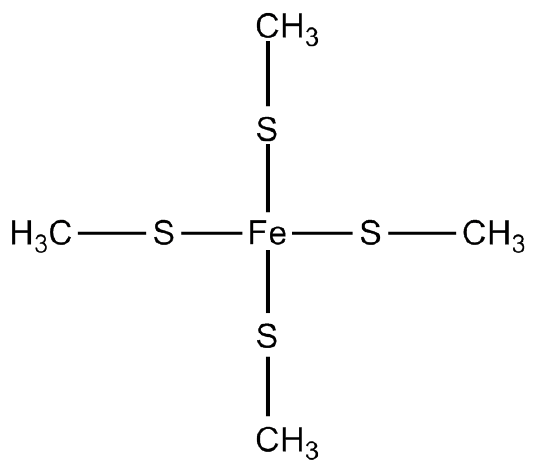}
\caption{Model system used to obtain the force constant parameters of the Fe-S bonds. }\label{fig:iron}
\label{fig:model}
\end{figure}

Because of the unavailability of parameters for the iron site, we build a bonded model whose initial structure is the crystal structure of mut-RbPa with the four C$_{\alpha}$ atoms of the cysteines replaced by hydrogens  (Fig.~\ref{fig:iron}).  To determine the equilibrium bond length of Fe-S, this model system is optimized using Gaussian 03~\cite{g03} with B3LYP exchange-correlation functional~\cite{becke:1993,lee:1988} and 6-31+g(d)/ 6-31++g(d,p)/ 6-311++g(d,p)/ 6-311++g(2d,2p) basis sets. The predicted distances conform to the crystal structure.
%~\textbf{Diana, why cite all these basis sets if we only use the last one? Please find out.}. 
The harmonic bond stretching force constant is calculated using the scheme presented in Ref.~\citenum{seminario:1996}. The frequency analysis is performed using Gaussian 03 with the same functional and basis set used to generate the Hessian of the optimized structure in Cartesian coordinates. %\textbf{Diana, I don't understand this last sentence.} 
For each Fe-S bond, a $3\times 3$ matrix 
\[\left[
\begin{array}{ccc}
\frac{\partial^2 E}{\partial x_{\mathrm{Fe}}\partial x_{\mathrm{S}}}&\frac{\partial^2 E}{\partial x_{\mathrm{Fe}}\partial y_{\mathrm{S}}}&\frac{\partial^2 E}{\partial x_{\mathrm{Fe}}\partial z_{\mathrm{S}}}\\
\frac{\partial^2 E}{\partial y_{\mathrm{Fe}}\partial x_{\mathrm{S}}}&\frac{\partial^2 E}{\partial y_{\mathrm{Fe}}\partial y_{\mathrm{S}}}&\frac{\partial^2 E}{\partial y_{\mathrm{Fe}}\partial z_{\mathrm{S}}}\\
\frac{\partial^2 E}{\partial z_{\mathrm{Fe}}\partial x_{\mathrm{S}}}&\frac{\partial^2 E}{\partial z_{\mathrm{Fe}}\partial y_{\mathrm{S}}}&\frac{\partial^2 E}{\partial z_{\mathrm{Fe}}\partial z_{\mathrm{S}}}
\end{array}
\right]\]
is extracted from the Hessian. Its eigenvalues $\lambda_i$ and eigenvectors $\mathbf{v_i}$ are obtained through diagonalization. The harmonic bond stretching force constant is then calculated as
\begin{equation}
k_{Fe-S}=\frac{1}{2}\sum_{i=1}^3\lambda_i | \mathbf{u}\cdot \mathbf{v_i}|,
\end{equation}
where $\mathbf{u}$ is the normalized vector linking the iron center to the sulfur. The coefficient $\frac{1}{2}$ conforms to the functional form of the Amber force field. The parameters for the oxidized and the reduced form of the site using the different basis sets are reported in Table~\ref{table:QM}. The small changes in the resulting parameters across the different basis sets indicate the robustness of the model.  For the purpose of this study, we use the Fe(III)-S 6-311++g(2d,2p) model parameterization. 
%Note that the equilibrium distance conforms to the crystal structure and that the large basis set guarantees a high precision for the calculation.

\begin{table}[h]
\small
\centering
\caption{Parameters for the iron site determined using ab initio calculations}\label{table:QM}
\begin{tabular}{cccc}
\hline
\multirow{2}{*}{\textbf{Bond type}}&\multirow{2}{*}{\textbf{Basis set}}&\textbf{Bond}&\textbf{Force constant}\\
&&\textbf{length (\AA )}&\textbf{(kcal mol$^{-1} \AA ^{-2}$)}\\
\hline
\multirow{4}{*}{Fe(III)-S}&6-31+g(d)&2.317&81.33\\
&6-31++g(d,p)&2.317&81.35\\
&6-311++g(d,p)&2.325&79.56\\
&6-311++g(2d,2p)&2.322&79.68\\
\hline
\multirow{4}{*}{Fe(II)-S}&6-31+g(d)&2.416&44.02\\
&6-31++g(d,p)&2.415&43.82\\
&6-311++g(d,p)&2.425&43.57\\
&6-311++g(2d,2p)&2.419&42.90\\
\hline
\end{tabular}
\end{table}

We verify the agreement between the PDB structure and the electron density map with MolProbity~\cite{Chen:2010}. For wt-RbPf, a clash was detected and the rotamer of Glu49 was accordingly adjusted.
This PDB structure is then immersed in water and ions, which neutralize the protein charge and recreate the high-salt experimental crystallization conditions. Steepest descent energy minimization and 100 ps-long simulations at constant volume $V$ and temperature $T$=300 K (constant NVT ensemble) with position restraints on the heavy atoms follow, in order to equilibrate the system temperature and relax the solvent. 

Because the proteins in these study have been previously crystallized, the crystal contacts, i.e. the regions on the protein surface that are within 4 \AA\ from each other in the observed protein crystal, are unique and easily identified from the PDB structure. We follow the soft matter hypothesis that these regions trigger crystal formation, so they correspond to the attractive patches in the coarse-grained model described in the following section. This hypothesis is validated by the rest of our analysis.

\begin{figure}[htb]
\centering
\includegraphics[width=0.45\textwidth]{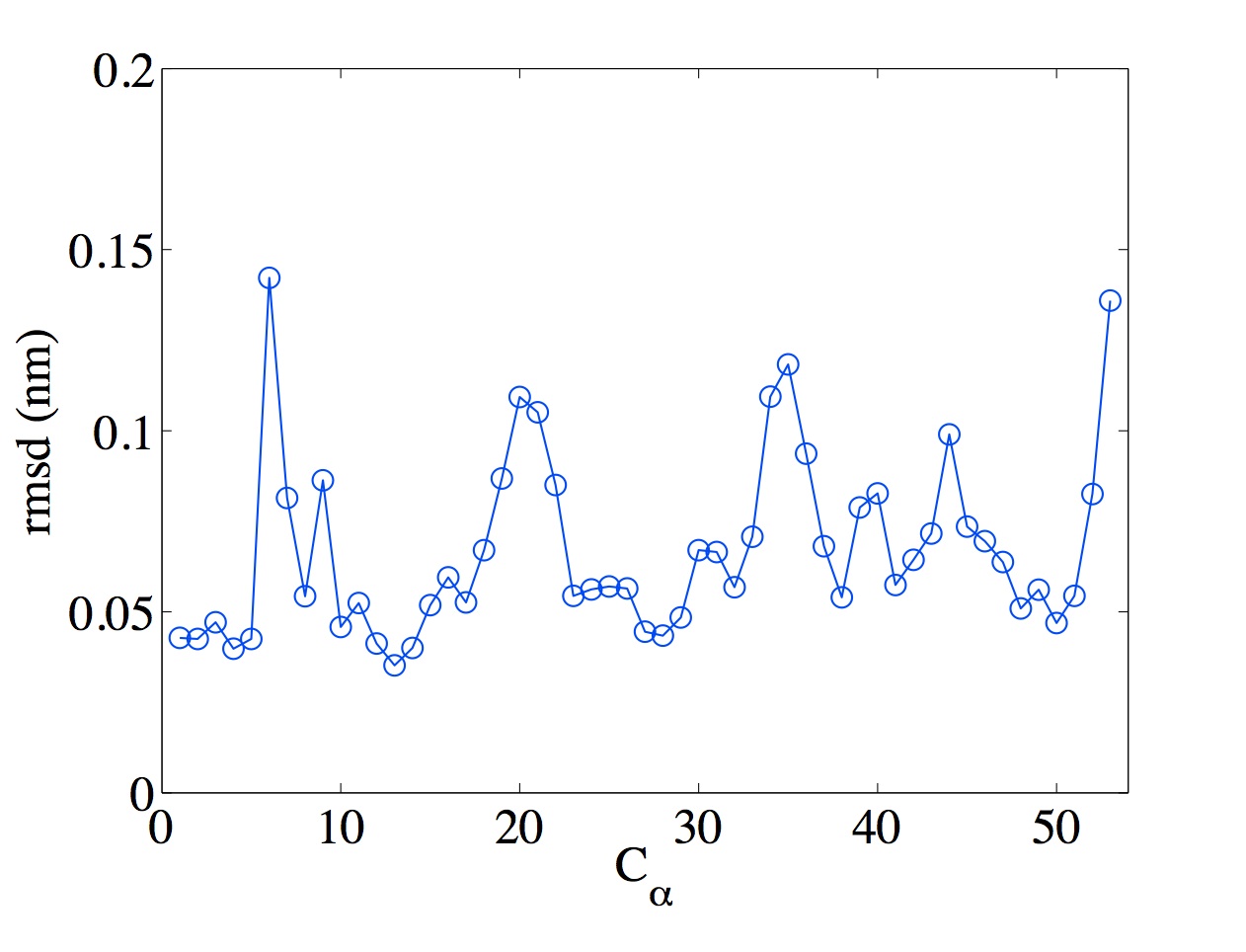}
\caption{Root mean-sqaured displacement (rmsd) fluctuations of the C$_\alpha$s along the umbrella sampling simulations. The regions of higher flexibility correspond to the protein's loops (around residues 7, 20, 35 and 42) and C-terminus, which are in qualitative agreement with earlier NMR studies~\cite{Blake1992}.}\label{fig:rmsf}
\label{fig:model}
\end{figure}

The PMF of each protein's every crystal contact is determined using the umbrella sampling and weighted histogram package implemented in Gromacs~\cite{Hub2010}. The principal axes of the inertia tensor of the interface provide two axes in the plane of the interface, and the third, which is orthogonal to the interface, is used as reaction coordinate. Starting configurations are generated by pulling the center of mass of one protein along the reaction coordinate while keeping the other protein fixed. 
During the pulling, we control the reciprocal orientation of the two proteins as in Ref.~\citenum{Pellicane2008}. We first determine the four most stable heavy atoms in the structure from individual MD simulations, such that the tetrahedron defined by these four centers spans most of the protein structure. We use the  $C_\alpha$ of Ile7, Pro19, Asp35 and Glu52 for wt-RbPf and mut-RbPa. For mut-RbPf, we use the $C_{\alpha}$ of Lys50 instead of Glu52, because the coordinates of the latter residue are not reported in the PDB file. We then restrain the angles between the edges of the two tetrahedra with a spring constant of 5000 kJ/(mol rad$^2$). These restraints prevent the proteins from rotating with respect to each other, yet allow ample freedom for both the flexible elements of the protein backbone and the sidechains to fluctuate, as shown by the typical fluctuation range of the C$_\alpha$s in the loops and C-terminus of the protein~\cite{Lindorff:2005} (Fig.~\ref{fig:rmsf}).
% \textbf{Should we reintroduce an updated figure about fluctuations (former text commented below), or is this not needed?}. 
The starting configurations are sampled every 1 \AA \ up to a distance of 10 \AA \ with a spring constant of 5000 kJ/(mol nm$^2$). A production run of 40 ns follows a 10 ns equilibration at constant temperature (300 K) and pressure (1 bar).

We estimate the umbrella sampling accuracy from 100 bootstraps~\cite{Hub2010}. The standard deviation associated with the PMF minimum is 2--3 kJ/mol, which corresponds to a 10-20\% relative error on the interaction strength. Using three replicates of the simulations provides a similar estimate. The uncertainty is comparable with the systematic errors introduced by a given choice of forcefield~\cite{Lindorff:2012}, and is thus in line with the overall robustness of the schematic model parameters.

The angular component of the interaction is obtained from four independent 5 ns-long simulations that restrain the distance between the centers of mass of the two proteins for each interface using a spring constant of 1000 kJ/(mol nm$^2$). 
To examine the role of salt in Sec.~\ref{sec:salt}, we perform 4 ns-long MD simulations during which the proteins' center of mass and reciprocal orientations are constrained to their crystal form. The ion distribution around the interface is measured every 20 ps. Their number density is obtained by normalizing over the available volume defined as the total volume minus the volume occupied by two spheres of diameter $\sigma$ centered at the proteins center of mass.

\subsection{Patchy particle model.}\label{sec:model}

In order to estimate the phase diagram of each protein and to identify their facile crystallization regime, we model proteins as hard spheres decorated by attractive patches representing the crystal contacts (Fig.~\ref{fig:model}). Each particle carries $i=1,\ldots,n$ pairs of patches. Patch  $2i$ interacts only with $2i-1$, as in the Sear model~\cite{Sear1999}, while the range and width of the interactions are independent parameters, as in the Kern-Frenkel model~\cite{Kern2003}.  In contrast to the Sear and the Kern-Frenkel models, which both assume the same interaction form for each pair of patches, we allow the interaction to vary from one pair of patches to another, capturing the chemically heterogeneous nature of the crystal contact interactions~\cite{dorsaz:2012}.  

\begin{figure}
\centering
\includegraphics[width=0.5\textwidth]{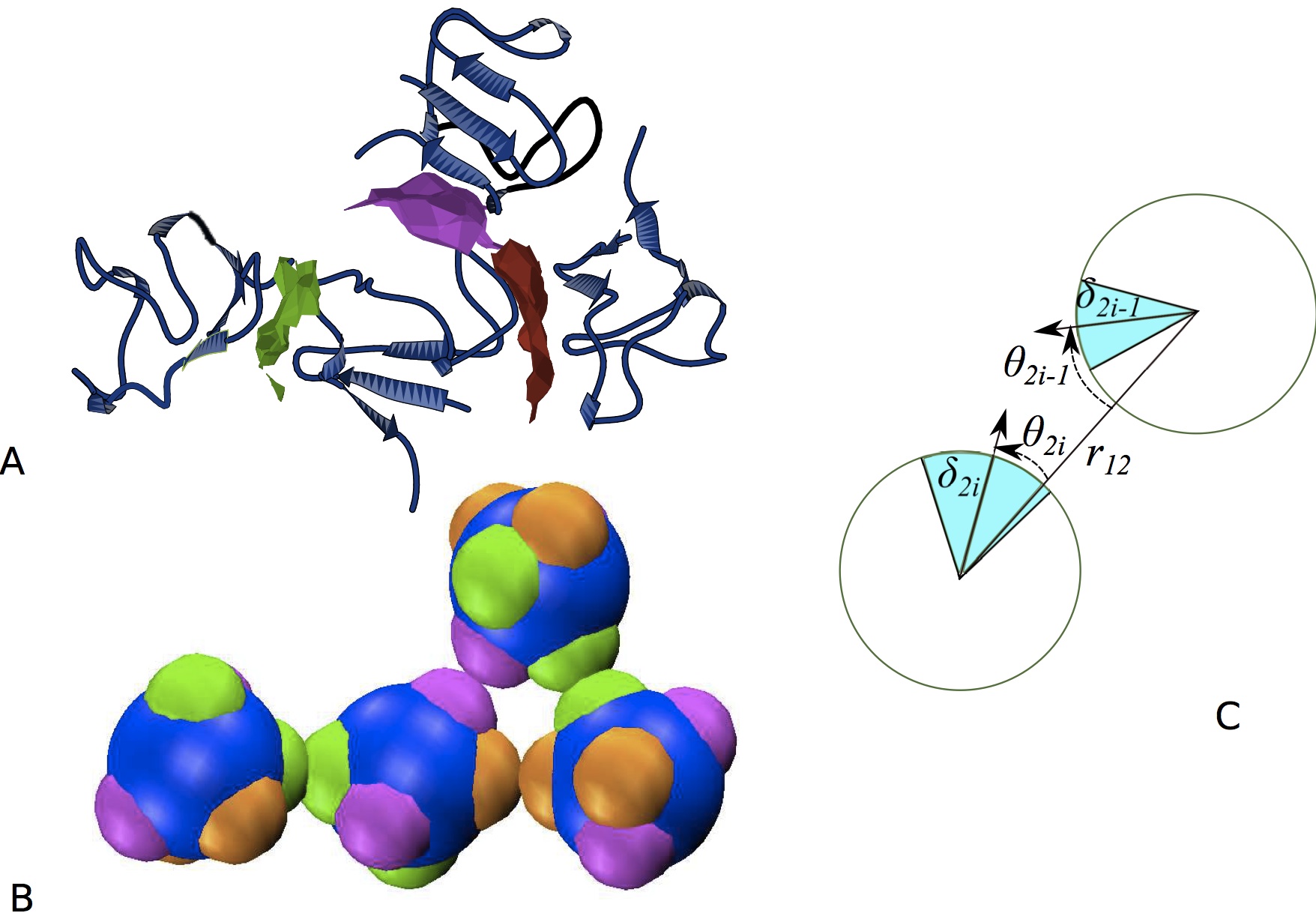}
\caption{Coarse-grained representation of the protein crystal (A) through a patchy particle model (B). The blue spheres are proteins on which each pair of patches corresponds to the crystal interface of the same color. C: schematic of the interaction between two particles through patch $2i$ and patch $2i-1$. The colored region represents the angular width of the patch. }
\label{fig:model}
\end{figure}

\begin{table}[htb]
\small
\centering
\caption{Model parameters for wt-RbPf, wt-RbPf at 45 mM of NaCl (low salt), mut-RbPf and mut-RbPa}\label{table:MC}
\begin{tabular}{ccccc}
\hline
\multirow{2}{*}{-}&\multirow{2}{*}{\textbf{wt-RbPf}}&\textbf{wt-RbPf}&\multirow{2}{*}{\textbf{mut-RbPf}}&\multirow{2}{*}{\textbf{mut-RbPa}}\\
&&\textbf{(low salt)}&&\\
\hline
$\sigma (nm)$&2.2&2.2&2.2&1.9\\
$\lambda_1$ ($\sigma$)&1.09&1.1&1.09&1.15\\
$\epsilon_1$ ($k_BT$)&3.7&0.6&3.7&8.4\\
$\cos\delta_{1}$&0.9&0.89&0.9&0.93\\
$\cos\delta_{2}$&0.89&0.89&0.89&0.89\\
$\lambda_2$ ($\sigma$)&1.1&1.1&1.1&1.1\\
$\epsilon_2$ ($k_BT$)&2.8&2.4&4.7&7.6\\
$\cos\delta_3$&0.95&0.95&0.95&0.95\\
$\cos\delta_4$&0.92&0.92&0.95&0.92\\
$\lambda_3$ ($\sigma$)&1.15&1.15&1.15&-\\
$\epsilon_3$ ($k_BT$)&3.3&2.9&3.3&-\\
$\cos\delta_5$&0.89&0.89&0.89&-\\
$\cos\delta_6$&0.89&0.89&0.89&-\\
\hline
\end{tabular}
\end{table}

The interaction between particles 1 and 2, whose centers are a distance $r_{12}$ apart, is thus
\begin{eqnarray}
\phi(r_{12},\Omega_1,\Omega_2)=&\phi_{\mathrm{HS}}(r_{12})+\sum _{i=1}^n[\phi_{2i,2i-1}(r_{12},\Omega_1,\Omega_2)\nonumber\\
&+\phi_{2i-1,2i}(r_{12},\Omega_1,\Omega_2)],
\end{eqnarray}
where $\Omega_1$ and $\Omega_2$ are the Euler angles. A hard-sphere (HS) potential captures the volume exclusion
\begin{equation}
\phi_{\mathrm{HS}}=\left\{
\begin{array}{cc}
\infty&r\leq\sigma\\
0&r>\sigma,
\end{array}\right .
\end{equation}
where $\sigma$ is the diameter of the particle. The patch-patch interaction is the product of a radial and an angular component
 \begin{equation}
 \phi_{2i,2i-1}(r_{12},\Omega_1,\Omega_2)=\psi_{2i,2i-1}(r_{12})\omega_{2i,2i-1}(\Omega_1,\Omega_2),
 \end{equation}
 where
\begin{equation}
\psi_{2i,2i-1}=\left\{
\begin{array}{cc}
-\epsilon_i&r\leq\lambda_i\\
0&r>\lambda_i
\end{array}\right .
\end{equation}
and
\begin{equation}
\omega_{2i,2i-1}(\Omega_1,\Omega_2)=\left\{
\begin{array}{cc}
1&\theta_{1,2i}\leq\delta_{2i} \hbox{ and } \theta_{2,2i-1}\leq\delta_{2i-1}\\
0&\hbox{otherwise}
\end{array}\right ..
\end{equation}
The interaction range between patch $2i$ and patch $2i-1$ is $\lambda_i$, $\delta_{2i}$ is the semi-width of  patch $2i$, and $\theta_{1,2i}$ is the angle between the vector $r_{12}$  and the vector defining patch $2i$ on particle $1$, as illustrated in Fig.~\ref{fig:model}C. An analogous definition holds for $\theta_{2,2i-1}$.  Following the soft matter convention, the model parameters are expressed in reduced units~\cite{frenkel:2001}. Lengths are in units of the sphere diameter $\sigma$, energies are in units of 300K$k_B$ ($\sim$ 2.49 kJ/mol), where $k_B$ is Boltzmann's constant, and temperatures are in units of 300K.

The model parameters are \emph{fully} determined from the MD simulations described above (Table~\ref{table:MC}). The particle diameter $\sigma$ is set by the range of the strong repulsion between proteins, the depth of the well $\epsilon_i$ corresponds to the PMF minimum, and the effective interaction range $\lambda_i$ is such that the specific patch contribution to the second virial coefficient matches that of the actual radial profile of the protein-protein interaction at the crystal contact. The angular distribution of the configurational space sets the width of the patches defined as
$\delta_{2i}=\min[\textrm{asin}(\frac{1}{2\lambda_i}),\textrm{acos}(1-2\tilde{\sigma}_{2i})]$, where $\tilde{\sigma}_{2i}$ is the standard deviation of $\cos(\delta_{2i})$ distribution from the MD simulations. The constraint $\sin(\delta_{2i})<(2\lambda_i)^{-1}$ guarantees that no patch can form more than one bond.
The location of the patches on the sphere is obtained from the crystal contacts on the protein (Fig.~\ref{fig:model}A and B). The protein crystal recorded in the PDB file is expanded using PyMol. The centers of mass of each protein in the unit cell are identified. The vectors connecting one protein's center of mass to its neighbors are then computed (the cartesian coordinates of the patches are reported in Table~\ref{table:coor}). Because of the P$2_12_12_1$ symmetry of the protein crystals in this study, each protein has six neighbors. In the case of mut-RbPa, only four neighbors are sufficiently close (surface distance less than 4 \AA) to be considered to be interacting, which is why only four patches are reported.

 \begin{table*}[htb]
\centering
\caption{Summary of the patches position and the crystal properties for the proteins in this study. The first section reports the cartesian coordinates of the normalized vectors of the models patches. Patch 1 interacts with 2, patch 3 with 4 and patch 5 with 6. The second section describes the crystal unit cell: unit cell dimensions and positions of the proteins center of mass within the unit cell. Both crystals have 4 proteins in their unit cell obtained by applying a two-fold screw axis symmetry along each direction to the reference protein in the asymmetric unit cell. The unit cell properties are obtained from the PDB file  }
\begin{tabular}{ccccccc}
\hline
\multicolumn{1}{c}{}
&\multicolumn{3}{c}{\textbf{wt-RbPf and mut-RbPf}}
&\multicolumn{3}{c}{\textbf{mut-RbPa}}\\
\hline
\hline
patch&x&y&z&x&y&z\\
\hline
1&-0.7191&0.5659&-0.4032&0.6268&-0.2698&-0.731\\
2&0.7191&0.5659&-0.4032&-0.6268&-0.2698&-0.731\\
3&-0.3813&-0.1669&-0.9093&-0.008&-0.9288&0.3706\\
4&-0.3813&-0.1669&0.9093&-0.008&0.9288&0.3706\\
5&0.3475&-0.7668&0.5397&&&\\
6&0.3475&0.7668&0.5397&&&\\
\hline
\hline
unit cell ($\sigma$)&1.551&1.585&1.986&1.257&1.975&2.254\\
\hline
position within the unit cell ($\sigma$)&&&&&&\\
\hline
reference&0&0&0&0&0&0\\
screw axis along x&0.775&0.610&1.551&0.629&1.705&1.521\\
screw axis along y&0.359&0.793&0.558&1.249&0.988&0.394\\
screw axis along z&1.135&1.403&0.993&0.620&0.727&1.127\\
\hline
\end{tabular}\label{table:coor}
\end{table*}
%Assuming that configurations around a PDB crystal contact are visited according to their Boltzmann weight transforms the ensemble of sampled angles into an estimate of the pair-wise interaction width. %\textbf{SHouldn't this sentence appear in the following subsection along with the rest of the model mapping?} 
%More specifically, given the model notation (see Sec.~\ref{sec:model}), we define $\delta_{2i}=\min[\textrm{asin}(\frac{1}{2\lambda_i}),\textrm{acos}(1-2\tilde{\sigma}_{2i})]$, where $\tilde{\sigma}_{2i}$ is the standard deviation of $\cos(\delta_{2i})$ \textbf{Boltzmann?} distribution \textbf{detail the transformation from angle distribution to Boltzmann weight.} obtained from these simulations.

The gas-liquid line of the phase diagram is obtained using the Gibbs ensemble method \cite{Gibbs} and the critical temperature and density using the law of rectilinear diameters~\cite{frenkel:2001}. The solubility line is computed by integrating the Clausius-Clapeyron equation starting from a first coexistence point, determined using free energy calculations and thermodynamic integration \cite{frenkel:2001,Vega2008} (the methodological details are the same as in Ref.~\citenum{fusco:2013}). 

\section{Results}\label{sect:results}

We describe below the MD results for the three proteins' crystal contacts and compare the experimental crystallization conditions with the protein phase diagrams obtained by parameterizing the patchy particle model with the MD results. We then analyze how salt affects the protein interactions and phase diagrams, in order to understand the high salt conditions used for crystallizing rubredoxin.

\subsection{Crystal contacts of wt-RbPf and mut-RbPf.}

\begin{figure*}[thb]
\begin{center}
\includegraphics[width=0.9\textwidth]{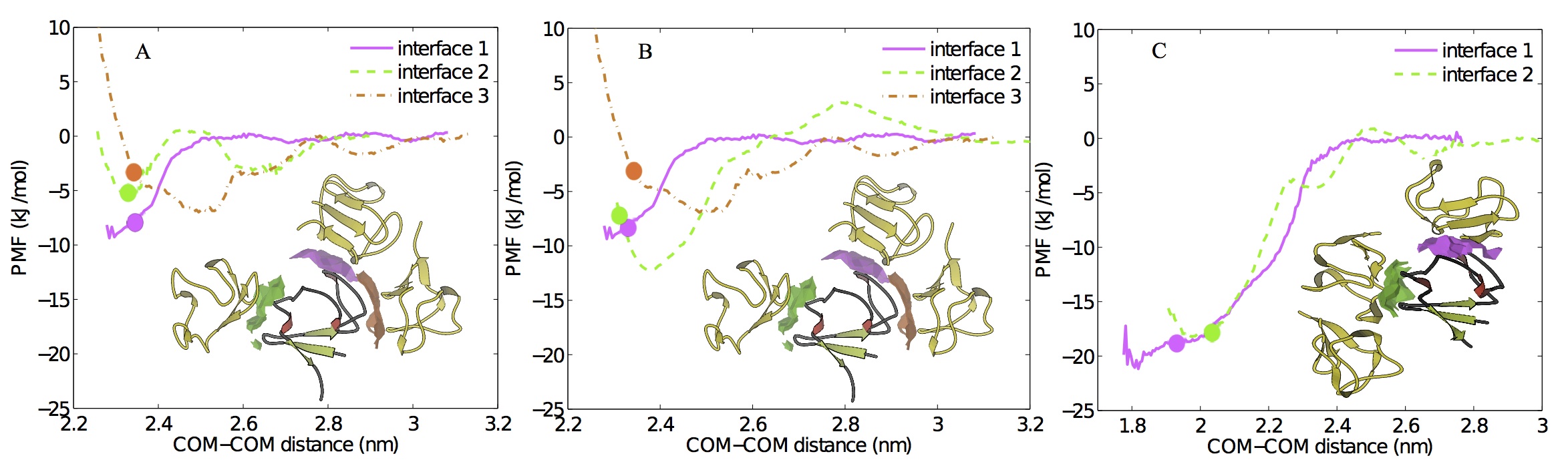}
\caption{PMF for wt-RbPf (A), mut-RbPf (B) and mut-RbPa (C). The filled circles indicate the COM-COM distance in the crystal. The interface surfaces between neighboring proteins identify the pair-wise interactions of residues involved in crystal contacts~\cite{Ban2006}.}
\label{fig:pmf}
\end{center}
\end{figure*}

Figure~\ref{fig:pmf} reports the solvated protein-protein interaction as a function of the center of mass-center of mass (COM-COM) distance and of the proteins' relative orientation for each crystal contact (1, 2, and 3) obtained using MD simulations. 
The orient\-at\-ionally-constrained potential of mean force (PMF) of wt-RbPf is first obtained in 3 M NaCl aqueous solution (Fig.~\ref{fig:pmf}A). 
The PMF minimum for interface 1a and 2a matches the observed crystal distance, while the crystal distance of interface 3a, although not exactly at the minimum of the PMF, is still within the attractive well.  The interfaces are differently attractive and show a varying range of interactions. The strong interface 1a contains a combination of negatively and positively charged residues suggesting that specific complementary amino acids come in contact and trigger the interaction. We will explore this interface in more details in Sec.~\ref{sec:salt}.
The relative abundance of apolar residues~\cite{Black1991,Headd2007}  in interfaces 2a and 3a compared to interface 1a suggests that hydrophobicity here drives the attraction. The major difference between these two interfaces is the presence of a barrier for interface 2a, which is likely caused by the interference of the disordered C-terminus of one chain with the crystal packing. The high B-factor of this region reported in the PDB file suggests it is highly flexible. The C-terminus appears to interact with the neighboring chain by forming a series of backbone-backbone hydrogen bonds at crystal distance (first minimum). When the two chains are at 2.5 nm apart the connection breaks. The C-terminus becomes very mobile and appears to impede the protein interaction, which is the source of the free-energy barrier. This oscillatory behavior in the PMF is thus likely unphysical and due to a combination of a poor choice of reaction coordinate and the slow dynamics of the C-terminus. Although in most cases constraining the movement along a single direction that is orthogonal to the interface plane is a reasonable approach for separating a pair of proteins, here this setup effectively limits the energetically favorable configurations between the C-terminus and the other chain by reducing the directions of approach between the two chains. In addition, the attachment-detachment dynamics of the tail at intermediate distance is much slower than the simulation time, which causes a poor equilibration of these simulations. As a result, the PMF profile of this interface is less reliable than the others, although it remains sufficiently accurate to parameterize the patchy model.

The effect of the C-terminus tail is also evident in mut-RbPf PMF (Fig.~\ref{fig:pmf}B). The mutations of this protein are not involved in any crystal contact, which leaves the crystal packing unchanged. The PDB file (PDB code: 1IU5), however, does not specify coordinates for the C-terminus tail, because the electron densities of these two residues is too poor for accurate modeling~\cite{Chatake2004}. We therefore compute the PMF for interface 2b  (identical to 2a) omitting the C-terminus (deletion of Glu52 and Aps53). The resulting profile does not show any significant barrier and reaches a more negative free energy. This finding supports the negative effect of the disordered C-terminus on crystal packing in agreement with the general understanding that highly-flexible regions of the protein may hinder crystallization.

The N-terminal methionine and formylmethionine variants (PDB codes: 1BQ8, 1BQ9)~\cite{Bau1998}  were crystallized under identical experimental conditions and show identical crystal forms, because the mutated region is not involved in any crystal contact.

\begin{figure}[hbt]
\centering
\includegraphics[width=0.4\textwidth]{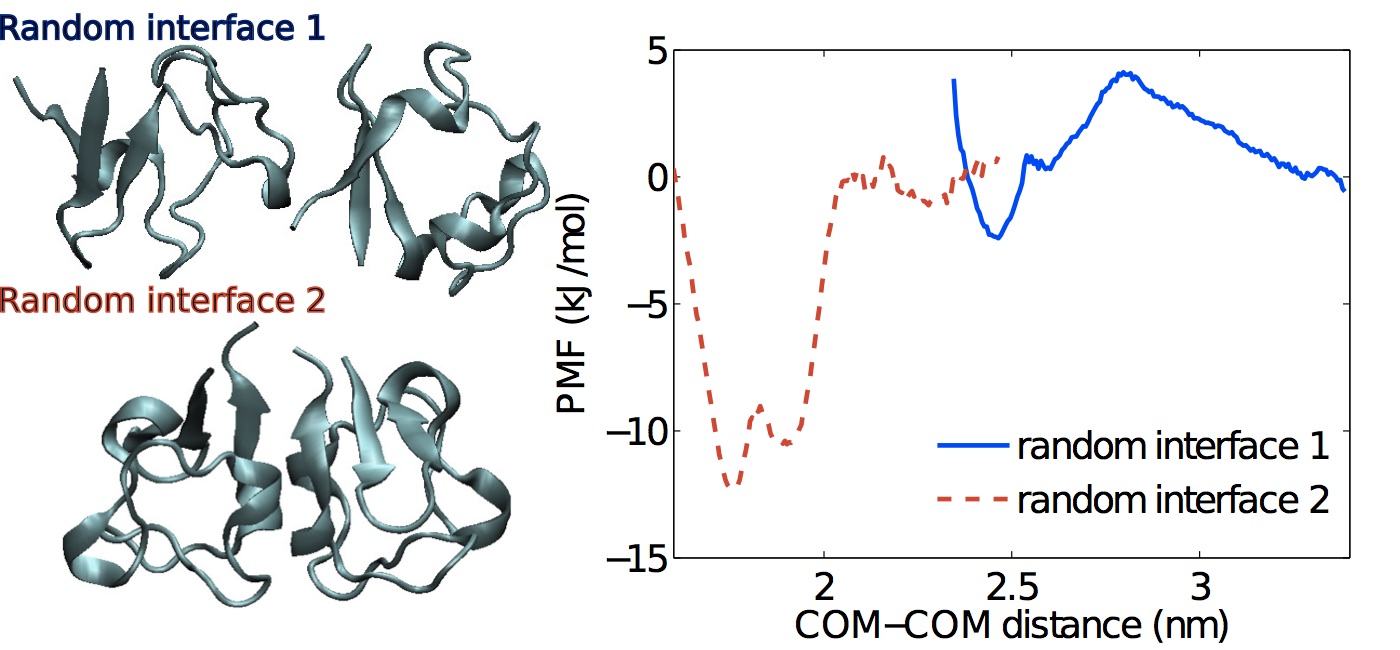}
\caption{PMF of two alternate protein-protein interfaces. Random interface 1 is built using one chain oriented as in interface 1a and one chain oriented as in interface 2a (solid line). Random interface 2 is the most favorable configuration found by RosettaDock (dashed line).}\label{random}
\end{figure}

This analysis of crystal contacts is not necessarily exhaustive if we assume that non-specific interactions can stem from hundreds of  thousands of protein-protein orientations~\cite{schmit:2012}. Yet these interactions have to be sufficiently strong compared to $k_BT$ in order to actually drive protein assembly. This threshold dramatically reduces the number of potential crystal contacts, and thus relevant patches.
To assess the randomness of crystal contacts, we determine the PMF for three alternate protein orientations that one may assume to be amongst the most attractive. 
%~\textbf{What about the second most attractive as I suggested? This section is a bit weak as is. We can talk about it if needed.}. 
The first is a hybrid of interface 1a and 2a, which determines whether a given patch only interacts with its partner patch. Figure~\ref{random} shows that the first interaction is mostly repulsive and weakly attractive at short range, in support of our patch-specific assumption. The second and third are the top two scored configuration found by RosettaDock~\cite{Gray2003,Chaudhury2007}. The highest scored interaction is as attractive as actual crystal contacts, but would result in a crystal of dimers because the surface regions of the two interacting chains involved in the contact are the same.  MD simulations show that the solid angle spanned by this interaction is very narrow ($\cos\delta=0.99$), which suggests that this contact is much more orientationally specific than typical crystal contacts (patch widths in Table~\ref{table:MC}). In addition, the PDB shows no record of a rubredoxin dimer, which suggests that the remaining open surface of the dimer presents no patch sufficiently strong to further drive crystal assembly. Finally, the second strongest interaction predicted by Rosetta is both non-dimeric and non-attractive (not shown), and we expect other contacts to result in even weaker interactions. This finding supports the notion that crystal contacts, although biologically non-functional, are definitely not in all ways similar to random protein-protein interfaces. They are undeniably characterized by some level of chemical complementarity and specificity.

%Otherwise, the interfaces share little in common. Interface 2a is either repulsive or neutral, while 
%interfaces 1a and 3a are both attractive.
%The relative long radial range of interface 1a can be explained by the presence of a pair of hydrogen bonds between the flexible C-terminus tail of one chain and a loop on the other.
%Interfaces 1a and 3a also have a different radial range (3 vs 2.45 nm), which can be explained by the stabilization of the former by a pair of hydrogen bonds between the flexible C-terminus tail of one chain and a loop on the other.
%Interestingly, the PMF of other reasonably attractive orientations result in interfaces that are either inaccessible due to the presence of high intermediate entropic barriers or not as attractive(Supplementary Material, Fig.~S2). A random surface patch therefore cannot result in a crystal contact, nor are such contacts isotropically distributed.

%The core mutations of mut-RbPf also result in a similar crystal form. Because of the low resolution of the C-terminus tail in this protein, the last two residues (Glu52 and Asp53) were not included in the PDB file (1IU5). Instead of assigning them a position, we characterize the pair-wise interaction in the absence of the C-terminus, which allows us to assess whether its presence is necessary for crystal formation. Interface 1b (equivalent to 1a) is most affected by the tail's removal. Without it, the interface area is smaller and the range of attraction shorter, but the interaction is similarly attractive suggesting that the C-terminus tail only weakly contributes to the interaction. The high motility of the terminus, suggested by its poor resolution in the PDB file, supports this conclusion. As we will see in Section 3.3, the absence of the terminus does not affect crystal formation.
% and the interaction weaker by about 7 kJ/mol than in the wild-type (Fig.~\ref{fig:pmf}B). As we will see in Section 3.3, the difference strongly affects crystal formation. 

\subsection{Crystal contacts of mut-RbPa.}\label{mut-RbPa}
Mut-RbPa has crystal contacts that are all substantially different from those of the previous two proteins. 
Although each chain has six nearest neighbors, only four of them (forming two interfaces) are sufficiently close to contribute to the pair attraction (Fig.~\ref{fig:pmf}C). Interface 2c is mostly hydrophobic, but the hydrophobic component on interface 1c is relatively small. %\textbf{Check this last sentence.} 
The attraction is instead dominated by a salt-bridge between Arg50 and Asp35, which is stable between 1.9 and 2.1 nm, while at larger distances water fills the gap between the two proteins. The PMF therefore first plateaus, then rapidly increases (Fig.~\ref{fig:pmf}C). The resulting distance at crystal contact for interface 1c and 2c are congruent with the PMF minimum.

\begin{figure}[hbt]
\centering
\includegraphics[width=0.5\textwidth]{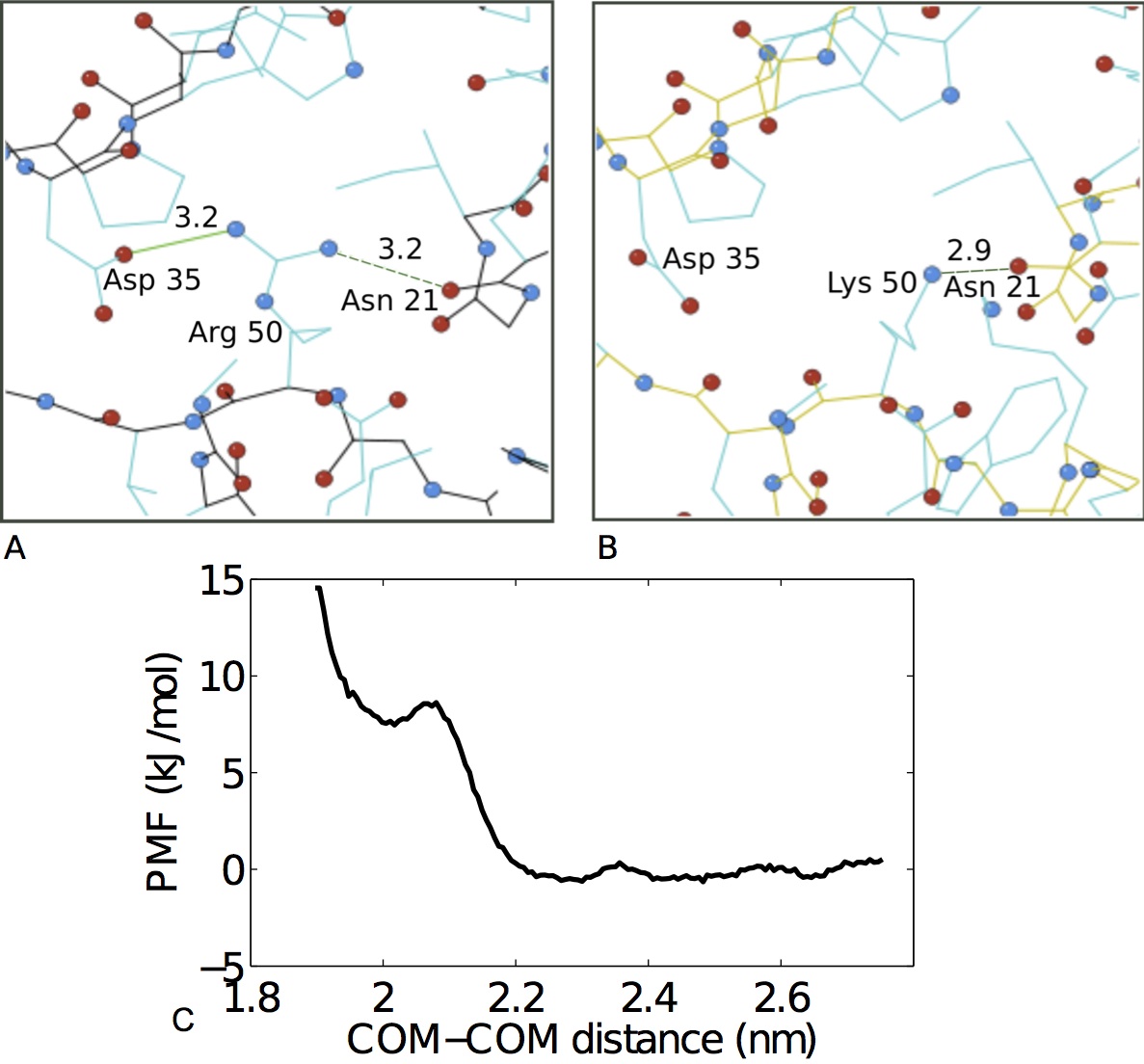}
\caption{A: Detail of the interaction between Arg50 and the neighboring chain in interface 1 in mut-RbPa. The solid line identifies the salt-bridge with Asp35 and the dashed line the hydrogen bond with the carbonyl group of Asn21. B: Interaction of Lys50 with the neighboring chain after deleting the last residue of wt-RbPf and fitting the structure in order to overlap the crystal contact of mut-RbPa. The lysine residue can either form the hydrogen bond or the salt-bridge, but not both. Bond lengths (in \AA ) are reported above the bond lines. C: The PMF for the interface represented in panel B indicates a neutral interaction between the two chains.}\label{magenta}
\end{figure}

Assuming that the strongest pair interaction determines the nucleation process, one may wonder why interface 1c in mut-RbPa is absent from the crystal contacts of wt-RbPf.
% ~\textbf{Do you answer this question? Should we reference something from later in the text?}
% In particular, interface 1c is at least twice as strong as any interface of wt-RbPf, while it is absent from the crystal contacts of wt-RbPf and its mutants.
We note that interface 1c misses the C-terminus and Lys50 is mutated to arginine. Deleting the C-terminus allows the two chains to fit closely together and the arginine residue to form a salt bridge with Asp35 (Fig.~\ref{magenta}A). To test whether the creation of this new interface can be attributed to the shorter tail of mut-RbPa, we delete the C-terminus of wt-RbPf and fit the two chains to the structure of interface 1c  (Fig.~\ref{magenta}B). The PMF indicates that the interaction between the two chains remains non-attractive (Fig.~\ref{magenta}C), i.e. lysine and arginine are not here interchangeable. Closer examination reveals that both Arg50 nitrogen groups interact  with the other chain through a salt bridge and a hydrogen bond. Conversely, lysine offers a single nitrogen group to compete with solvation. In agreement with lysine's solvation free energy being nearly twice more negative than that of arginine~\cite{Bash1987}, we find that lysine solvates immediately and does not interact with the other protein chain. This finding suggests that the lysine at position 50 in RbPf prevents the formation of a crystal contact analogous to interface 1c in mut-RbPa. Replacing this lysine with an arginine  should thus favor a 1c-like interface in RbPf and allow wt-RbPf and mut-RbPf to crystallize isomorphously to mut-RbPa.

\subsection{Patchy particle models and phase diagrams.}
 
\begin{figure}
\centering
\includegraphics[width=0.5\textwidth]{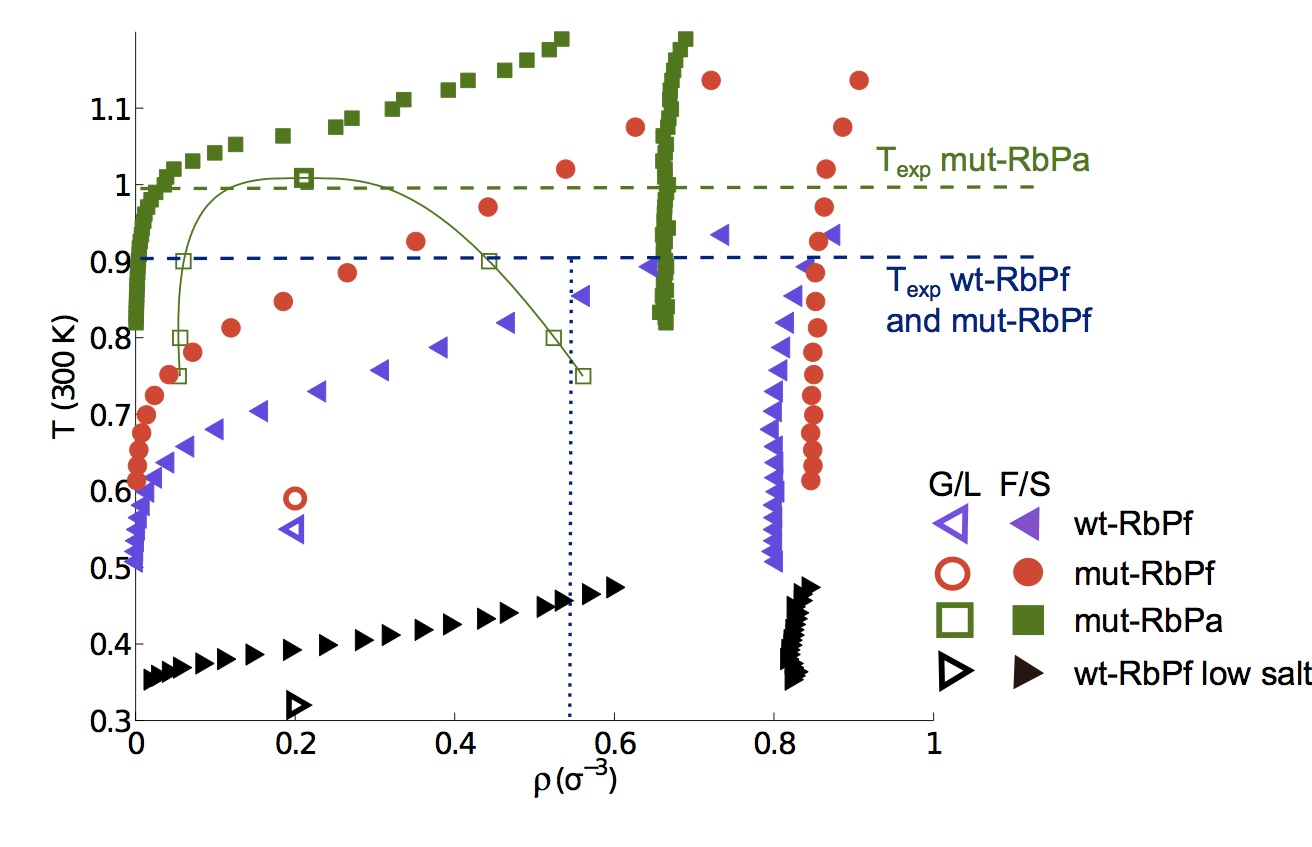}
\caption{Phase diagram for wt-RbPf, mut-RbPf and mut-RbPa at 3 M of NaCl and wt-RbPf at 45 mM of NaCl. The fluid-solid lines (F/S) for the four systems are represented respectively by left-pointing triangles, circles, squares and right-pointed triangles. The gas-liquid (G/L) are indicated by the corresponding empty symbols (thicker edge symbol indicates the critical point).  The dashed lines correspond to the temperatures at which the crystallization experiments were conducted. The dotted line indicates the maximum reachable density for the experimental protein solution of wt-RbPf.}
\label{fig:phase_diagram}
\end{figure}

Figure~\ref{fig:phase_diagram} illustrates the phase diagram for patchy particle models parameterized with the above MD results for each protein. 
The fluid-solid line (F/S) marks the coexistence conditions between solvated and crystallized proteins, while the gas-liquid coexistence line (G/L) identifies the conditions under which high and low concentration solutions of proteins can coexist. As expected from the relatively short radial range of the attraction ($\sim5-15\%\sigma$), in all cases the G/L line is metastable with respect to the solubility line.  
The observed relative variations of the attraction width (between 1 and 5\%) and range (between 5 and 10\%$ \sigma$) only change the melting temperature by up to 3\% and the critical temperature by up to 8\%, which supports the robustness of the schematic models in the crystallization zone. The phase diagrams allow us to predict the conditions under which crystallization can be successful, i.e., between the solubility line and the critical point, the conditions under which the protein will be undersaturated, i.e., above the F/S line, and the conditions under which over-nucleation and amorphous aggregation occurs, i.e., below the G/L line. 
These predictions can be straightforwardly validated by comparing these phase diagrams with the experimental crystallization conditions. Remarkably, for all three proteins the experimental temperature falls within 10\% of the crystallization gap, which is comparable with the estimated uncertainty of the coexistence lines due to the error in the MD simulations. 

In the case of wt-RbPf, it is not unreasonable that an unusual high density has to be reached to cross the solubility line. Given the initial experimental protein and salt concentrations in the sample and in the buffer, the protein concentration can reach up to $\rho=\frac{N}{V}\approx 0.55$\footnote[3]{We define $\rho_{\mathrm{p}}^i=\frac{N_{\mathrm{p}}}{V^i}$ as the initial protein density, $\rho_{\mathrm{salt}}^i=\frac{N_{\mathrm{salt}}}{V^i}$ as the initial salt density and $\rho_{\mathrm{salt}}^f=\frac{N_{\mathrm{salt}}}{V^f}$ as the final salt density assumed to be equal to the salt density in the buffer, then the final protein density is
\begin{equation}
\rho_{\mathrm{p}}^f=\frac{N_{\mathrm{p}}}{V^f}=\frac{N_{\mathrm{p}}}{N_{\mathrm{salt}}}\rho_{\mathrm{salt}}^f=\frac{N_{\mathrm{p}}}{N_{\mathrm{salt}}}\rho_{\mathrm{salt}}^f\frac{V^i}{V^i}=\frac{\rho_{\mathrm{p}}^i}{\rho_{\mathrm{salt}}^i}\rho_{\mathrm{salt}}^f. 
\end{equation}}.
This observation in conjunction with the higher error of the MD results for interface 2a rationalizes the more limited agreement between model and experiment in this case.

\begin{figure}
\centering
\includegraphics[width=0.5\textwidth]{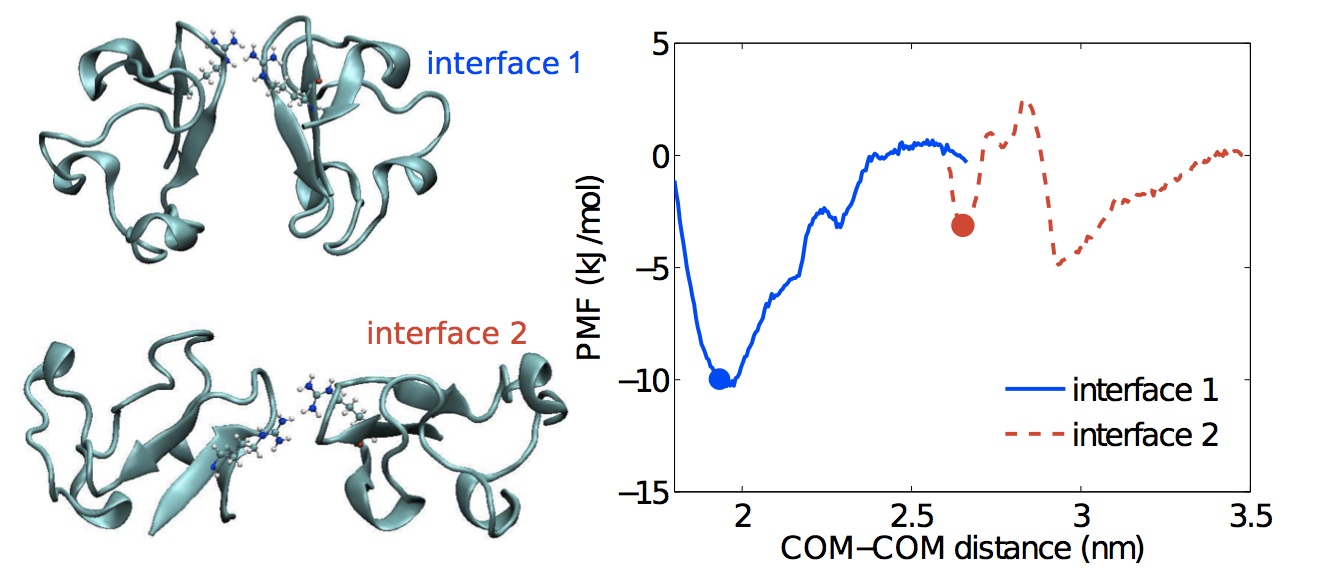}
\caption{PMF as a function of COM-COM distance for the two interfaces involving Arg5 (explicitly shown in the left panel) in wt-RbPa. The simulations, run at 3 M of NaCl and without dioxane, show much weaker attraction than the crystal contacts of mut-RbPa.}\label{1yk5}
\end{figure}

For the phase diagram of mut-RbPa, we find the temperature at which the protein was experimentally crystallized (293 K) to be very near the model's critical temperature, close to the amorphous regime. 
Interestingly, in the same crystallography study, the wild type of RbPa was crystallized at a poorer resolution and in a different unit cell by adding dioxane to the sample~\cite{Bonish2005}. By comparing the crystal forms of the two proteins, we identify one mutation (Arg5) that is involved in two crystal contacts of the wild-type, but not of the mutant, which can in principle trigger the formation of a different crystal form. Simulations show, however, that the contacts involving the mutation are much less attractive than the mut-RbPa crystal contacts in the absence of dioxane (Fig.~\ref{1yk5}). This result suggests that the wild-type should crystallize isomorphously to the mutant under the mutant's crystallization condition, i.e. without dioxane. Because similar crystal contacts should correspond to similar phase diagrams, we expect these solution conditions to be near the G/L line for the wild-type, as they were to the mutant's. It is therefore not surprising that the study reports excessive nucleation of the wild-type when crystallization was attempted without dioxane. Small perturbation of the wild type interactions due to the mutations or to variations in the experimental solution concentrations may have sufficed to tilt the system below the G/L line. 
Two experimental approaches can then overcome the problem: weakening the protein interactions by changing the solution conditions, or increasing the solution temperature. The authors of the study, as many before them, opted for the first approach and added dioxane to the solution.  This additive enhances electrostatic interactions and results in an effective repulsion between proteins, which (naturally) carry the same net charge.  Increasing the temperature might, however, have been a better strategy than adding a cosolute, in order to reproducibly obtain high-resolution crystals. Such strategy has already proven useful in obtaining better quality crystals in other proteins, but it is ``often neglected despite its proven impact"~\cite{Landsberg2006,Benvenuti2007}, possibly because of the poor microscopic understanding of the approach until now.

It is important to note that the assumption of similar phase diagrams only holds if the residues that differ between the two proteins do not strongly affect protein-protein interactions, which is not always the case. Comparing wt-RbPf and mut-RbPa shows that a single mutation (lysine to arginine) there dramatically affects the crystal organization (Sec.~\ref{mut-RbPa}).

%The identical experimental crystallization conditions for wt-RbPf and mut-RbPf, as well as for a series of mutants whose crystal contacts are closely related
%~\cite{Bau1998,Chatake2004} suggest that the phase diagram of the two proteins should be very similar. Removing the C-terminus tail, which weakens 1b, however, is found to lower the solubility line well below the experimental crystallization temperature. Although absent from the PDB structure, our analysis suggests that the C-terminus in the mutant should similarly interact with the other chain (as in the wild type) for crystallization to be physically possible. We cannot exclude that a mutant of RbPf missing the C-terminus could have the same crystal contacts as the wild type, but were it to be the case our study predicts that such a mutant would not crystallize under the same solution conditions. 
\subsection{Effect of low salt concentration on wt-RbPf.}\label{sec:salt}

All crystallized rubredoxins have been precipitated out of relatively high salt concentration ($\sim$3 M) solutions. In order to get a clearer understanding of this feature and to validate the robustness of our method, we also calculate the PMF of wt-RbPf at low salt concentration (45 mM). This approach enables us to identify the differences that make this solution condition unsuitable to crystallizing rubredoxin.

\begin{figure}[hbt]
\centering
\includegraphics[width=0.4\textwidth]{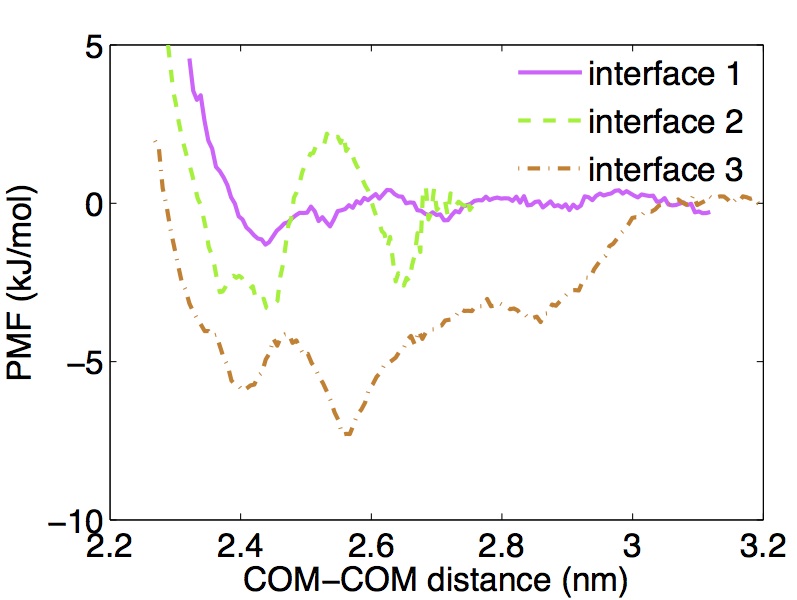}
\caption{PMF as a function of COM-COM distance for wt-RbPf in a 45 mM solution of NaCl.}\label{brf_low_salt}
\end{figure}
%Even taking the C-terminus contribution into account, the experimental crystallization temperature of RbPfs (277 K) is still far too high for our phase diagram. 
%A possible source of discrepancy is the different NaCl concentration used in the patchy model parameterization compared to the experimental conditions (45 mM vs.~$\sim$ 3 M). 

\begin{figure}[htb]
\centering
\includegraphics[width=0.4\textwidth]{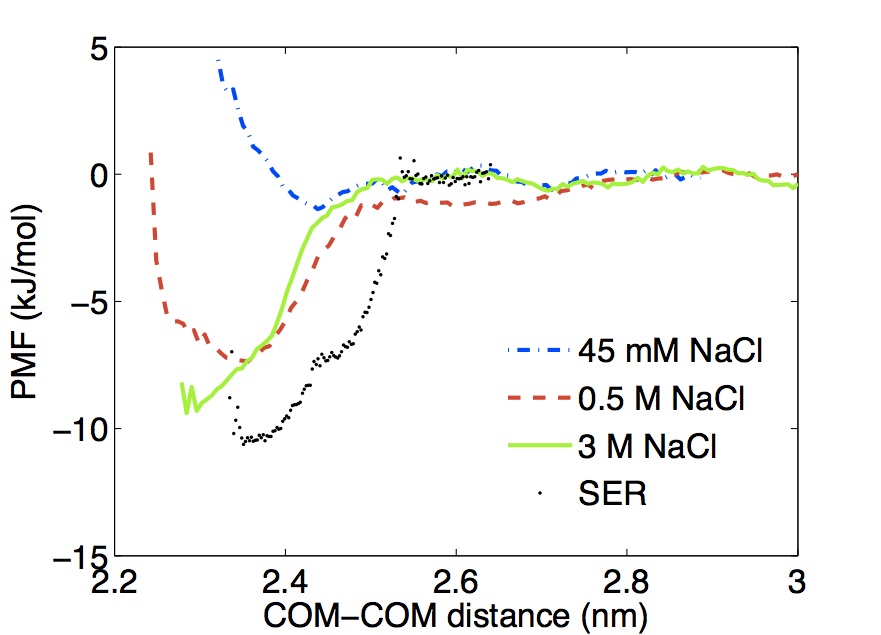}
\caption{PMF of interface 1 for wt-RbPf at different salt concentrations and for the E49A mutant at 45 mM of NaCl (SER).}
\label{fig:pmf_int1}
\end{figure}

Salt only weakly perturbs the PMF of interfaces dominated by hydrophobic attraction  (within the simulation error, Fig.~\ref{brf_low_salt}), in agreement with NaCl being a weak salting-out agent~\cite{Zangi2007}. Interface 1a, however, becomes significantly less attractive at 45 mM NaCl (Fig.~\ref{fig:pmf_int1}). The microscopic origin of this effect is detected by examining the behavior of ions around the interface at high and intermediate salt concentration (3 M and 0.5 M of NaCl). 

\begin{figure}
\centering
\includegraphics[width=0.5\textwidth]{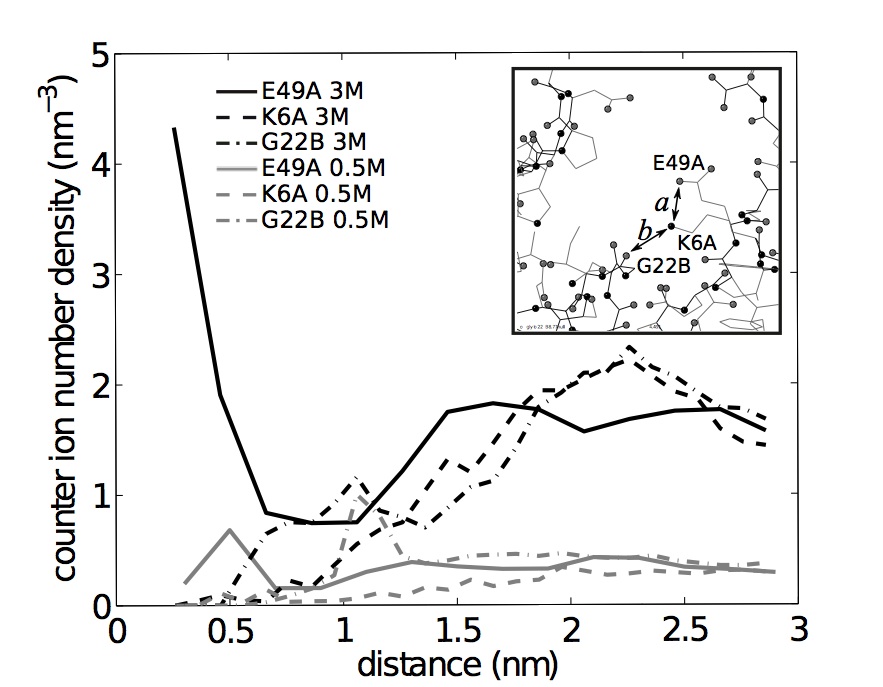}
\caption{Number density of counter ions as a function of distance from the charge of three different residues: Glu49 on chain A (solid), Lys6  on chain A (dashed) and the oxygen of Gly22 on chain B (point-dashed). Black lines refer to the 3 M NaCl simulations, grey lines to the 0.5 M NaCl simulations. The inset shows the location of the residues at crystal contact. At low salt concentration Glu49 interacts with Lys6 on the same chain (arrow $a$), while at higher salt concentration Glu49 is screened and Lys6 interacts with the carboxyl group of Gly22 on the other chain (arrow $b$). }
\label{fig:screening}
\end{figure}

Interface 1a contains six negatively (two on one chain and four on the other) and three positively  (all on a single chain) charged residues. It should thus be highly hydrated, as it is found in both the PDB structure (1BRF) and the simulations. The overall electrostatic repulsion between the two proteins at this interface is thus weakly screened at low salt concentration, but the situation is different at high salt concentration. The positively charged residues at the edge of the interface are then accessible to ions and therefore screened, while the negatively charged residues buried in the core of the interface are much less screened. As a result, like charge repulsions between residues on different proteins and opposite-charge attractions between residues on the same protein are weakened, while unlike charges on different proteins at the core of the interface remain essentially unscreened. A net effective attraction between the two chains is thus observed (Fig.~\ref{fig:screening}). In other words, when no ions are present, the positively charged Lys6 interacts with Glu49 on the same chain and not with the polar carboxyl of Gly22 on the other chain. If the salt concentration increases, Glu49 is entirely screened by counter ions beyond 2 \AA \  (black solid line), but both Lys6 and Gly22 still interact with each other. To further support this interpretation, we observe that by replacing Glu49 with a neutral alanine the resulting PMF is attractive even at low salt concentration (SER profile in Fig.~\ref{fig:pmf_int1}).

The phase diagram obtained by parameterizing the model's interface 1a at low salt concentration pushes the solubility line to lower temperature and moves the crystallization gap further away from the experimentally accessible temperature range. The model, therefore, suggests a nearly-quantitative explanation for choosing high salt crystallization conditions for this family of proteins.
%Reparameterizing the patchy model's interface 2a brings the phase diagram of wt-RbPf in remarkably close agreement with the experimentally observed crystallization conditions (Fig.~\ref{fig:phase_diagram}), which validates the overall approach.

\section{Discussion}\label{sect:discussion}

From this proof-of-concept study, we can draw microscopic insights about the underlying structural biology and soft matter motivations for its inception.

\subsection{Surface Entropy Reduction.}\label{SER}
The surface entropy reduction mutagenesis approach, rooted in extended crystallographic expertise~\cite{derewenda:2004}, has been statistically justified by mining the PDB~\cite{Derewenda2009}. 
Although its microscopic basis lies on reducing the entropic cost of crystallization, it also implicitly accounts for other effects by targeting specific residue types (lysines and glutamic acids) and not other equivalently highly entropic side chains (arginines and aspartic acids)~\cite{Goldschmidt2007}. 

The analysis of crystal contacts presented here allows us to go beyond the statistical justifications and offers a physical rationale for the observed difference in behavior between residues with a similar entropy.
In interface 1c of mut-RbPa, for instance, we verify that replacing Arg50 with a lysine significantly weakens the energetic interaction, in spite of the two residues having a similar entropy~\cite{Derewenda2009}. In this specific case, the different propensity of forming multiple intermolecular interactions and the solvation free energy play key roles in distinguishing the contribution of the two side chains. This effect thus offers an explanation as to why lysine is a good SER target, while arginine is not. 

We also obtain an example where a targeted mutation is advantageous from both a SER and an energetic point of view. We show that at low salt concentration wt-RbPf would not crystallize, or at least not in the same form. As observed in the analysis of interface 1a, the electrostatic interaction of Glu49 with Lys6 at low salt concentration prevents the inter-chain interaction between Lys6 and Gly22 and therefore also the protein's crystallization. For wt-RbPf, the SER approach specifically recommends replacing Glu49 with alanine~\cite{Goldschmidt2007}. Mutating Glu49 deletes the responsible charge and strengthens the pair interaction by an alternate route to increasing the salt concentration. In this situation, the suggested mutation would likely help crystal formation in part because it reduces the surface entropy, but mainly because it changes the interface electrostatic potential (Fig.~\ref{fig:pmf_int1}). 

The microscopic analysis of the crystal contacts in this work provides physical evidence that both support and modulate the statistical findings about them~\cite{Derewenda2009}. For one thing, compression of the lattice unit cells and readjustments of polar residues are common effects of crystal freezing~\cite{Juers:2001}, which can bias analyses based on PDB structures alone. Our solvated approach overrides some of these biases by simulating experimental crystallization temperatures, where the protein properties are averaged over a thermal ensemble of configurations. This clarified picture also suggests ways to improve the success of SER-like methods.
 Analyzing the nature of the surrounding amino acids along the chain, for instance, should help identify residues that weaken interactions by disrupting hydrophobic patches or by competing with favorable inter-chain interactions. 
 
 The general strategy of strengthening protein interactions does not, however, always results in better crystals. The connection we establish between protein-protein interactions and the overall protein-solution phase diagram clarifies this key point. If a protein gels or over nucleates, the crystallization conditions should instead be chosen to weaken these interactions, which can be achieved, for some maybe counterintuitively, by increasing temperature and thus entropy's contribution to protein-protein interactions.

\subsection{Patchy models.}
The central premise that the phase behavior of crystallizing proteins could be understood from that of particles with short-range anisotropic interactions is here verified, at least for a set of small and compact globular proteins. The typical regime for successful experimental crystallization is intermediate between the metastable critical point and the solubility line at low to intermediate protein concentrations, which results in open crystal structures dominated by directional interactions~\cite{dorsaz:2012}.

It is by now well understood that reducing the range and surface coverage of attraction lowers the critical temperature and broadens the crystallization regime \cite{Vega1992,Kern2003,Bianchi2011}. 
The critical point may or may not assist nucleation~\cite{Wolde1997,Haxton2012}, but the question is of little relevance if the protein is in any case highly soluble. Our study shows that in a low-salt aqueous environment RbPf and its mutants, for instance, do not crystallize because their solubility lines lie either at very low temperatures or at high densities. The addition of salt to the solution strengthens the pair interaction, but does not significantly affect its range or width, except for incompletely screened electrostatic interactions. The pair interactions therefore naturally results in an attraction range that  lies between 1 and 3 \AA. The parameter that most affects solubility is the interaction strength. The phase diagram in Fig.~\ref{fig:phase_diagram} shows that in order to achieve reasonable solubility at room temperature, the average interaction strength per patch should be of order 8 kJ/mol for a protein with six crystal contacts and of 20 kJ/mol for a protein with four.
A larger protein would shrink the fraction of the protein diameter $\sigma$ over which the attraction is felt as well as its angular span. Although tightening the fractional range by an order of magnitude dramatically affects the critical point location, it only weakly perturbs the position of the solubility line. If anything, crystallization may then become easier. The fact that it generally is not highlights the incompleteness of this description, which should then probably include a larger number of patches for larger proteins. We get back to this question in the conclusion.

A common assumption in patchy particle models is that the patch-patch interactions responsible for crystallization are essentially identical. Our simulations clearly show that it is not the case in proteins. Comparing the phase diagram of wt-RbPf and mut-RbPa indicates that it may be more efficient to have many weaker patches than few stronger patches, in order to decrease the solubility of a protein. If we rescale the temperature over the average energy per particle in the crystal ($\sim$8 $k_BT$ for wt-RbPf and $\sim$16 $k_BT$ for mut-RbPa), the solubility line of mut-RbPa drops below that of wt-RbPf, indicating that mut-RbPf is crystallized by making up for the absence of an interface by having a drastically increased attraction strength for the others. This observation motivates the study of how the distribution of energy across patches affects the phase diagram~\cite{fusco:2013}.

Modifying the interaction strength is equivalent to rescaling temperature. A patchy model with stronger interactions maps to the same phase diagram with higher temperatures. From a practical point of view, except for small changes to solution temperatures, tuning the strength of the interaction is, however, quite difficult to achieve and typically requires a detailed microscopic understanding of the system. Our results show that even small details, such as mutating a lysine to an arginine, or the preferential screening observed at high salt concentration, can significantly affect the interaction. Echoing the SER discussion, we thus urge for more systematic atomistic-level studies of this question. 

\section{Conclusion}\label{sect:conclusion}

In this paper, we have studied the interaction of three closely related proteins of the rubredoxin family using an approach that bridges the structural biology and soft matter descriptions of protein crystallization. It allowed us to characterize representative values of protein-protein interactions and to obtain reasonable phase diagrams, providing a microscopic explanation of why certain simple mutations can dramatically affect a protein's crystallization behavior. The correspondence between protein crystallization and patchy models phase diagrams provides guidelines to avoid gelation and over-nucleation and, more generally, draw stronger parallels between schematic models for soft matter and protein assembly.

Because the goal of this study is to verify the correspondence between patchy particle models and real proteins, and to identify microscopic mechanisms that trigger these interactions, we have here focused on a protein for which structure and crystal contacts are known. In this case the interacting patches are directly extracted from the known protein crystal contacts. Although one does not typically have access to such detailed information when attempting to crystallize a protein \textit{de novo}, this approach can nonetheless be of merit even when patch identification cannot be directly read off from a PDB file. For instance, if a low-resolution crystal is available, a rough estimate of the protein structure can be fit to the electron density map and the crystal contacts so analyzed. MD simulations would then allow to relax and correct a defective initial structure and give reasonable results for the crystal contact interactions. The phase diagram of the resulting model can be used to guide subsequent attempts at improving the crystal resolution. Similarly, if the structure of a homologue or a mutant is known, MD simulations of its crystal contacts in which the differences between the target protein and the one available are replaced \textit{in vitro} can be used to obtain an initial phase diagram and tune the conditions to obtain an isomorphous crystal for the target protein.

In the worst case scenario, the case in which nothing is known beyond  a protein's primary sequence, our approach has to be preceded by some analysis of the protein surface. Multiple iterations between simulations and experiments would then likely be necessary. Protein folding algorithms, although far from reliably and systematically predicting the full three-dimensional structure of a protein, may help guess which residues are likely to be on the surface and closed to which others~\cite{Dill:2012}. This rough surface map could then be used to identify problematic regions that hinder protein-protein interactions and thus determine good targets for mutations, as suggested in Section~\ref{SER}. 
Our microscopic analysis further suggests that crystal contacts are mostly characterized by specific patterns of hydrophobic, charged and polar residues and by the presence of few residues, i.e. arginine, that have peculiar chemistry. This observation justifies and encourages extensive simulations to characterize the interaction between patterns of these residues in different solution conditions. This information could be collected in a dictionary that is searched to find potential crystal contacts on a protein given a coarse map of its surface. The set of potential crystal contacts could then be used as set of patches in the model to obtain a first guess of phase diagram to tune crystallization conditions.
The location of the patches, although important in the actual protein and in the formation of a crystal versus another, only weakly affects crystallization in patchy models~\cite{fusco:2013} and can, therefore, be adjusted in a second step. We anticipate that future studies will clarify the usefulness of such a scheme.

Considering more complex proteins would also require introducing additional features to the schematic models. For example, some proteins assemble in more than one crystal form, each involving a distinct set of crystal contacts. Consequently, they should be characterized by a larger set of patches. Although different crystal forms might result in very similar protein structures, they are nevertheless interesting because they can produce higher resolution crystals or highlight dynamical features that would otherwise be unnoticed in the static crystallographic view, such as the hinge-bending angle in lysozyme~\cite{Zhang:1995}. The position and the parameterization of the patches may also not be kept fixed if conformational changes occur on a timescale similar to crystallization, such as in intrinsically disordered proteins. Work in this direction should help clarify both the soft matter and structural biology viewpoints.

\section*{Acknowledgments}
We thank Daan Frenkel, Dewey McCafferty, Terrence Oas, David and Jane Richardson, Weitao Yang and Wei Yang for various conversations. We acknowledge support from National Science Foundation Grant No. NSF DMR-1055586 and National Institute of Health Grant No. NIH GM-061870.

\bibliography{MD_paper} %your .bib file
\bibliographystyle{prsty} %the RSC's .bst file

\end{document}